\documentstyle[twocolumn,aps,prl,graphics]{revtex} 
\draft
\title{A Transient Instability in Thin Films of $n$CB Liquid Crystals}
\author{Stefan Schlagowski \and Karin Jacobs \and Stephan
  Herminghaus}
\address{Dept. of Applied Physics, Ulm University, D-89069 Ulm, Germany}

\begin{document}
\maketitle

\begin{abstract}
A transient surface instability is reported in thin nematic films of 
5CB and 8CB, occurring near the nematic--isotropic phase transition. 
Although this instability leads to patterns reminiscent of spinodal 
dewetting, we show that it is actually based on a nucleation mechanism. 
Its characteristic wavelength does not depend markedly on film thickness, 
but strongly on the heating rate.
\end{abstract}

Following several studies on the spreading behavior of liquid
crystals (LC) from the $n$CB homologous series \cite{BarS96,BarO99}, 
undulative instabilities have been observed in thin films of the LC 
5AB$_4$ \cite{HerJ98}, and 5CB \cite{VanV99} near the nematic--isotropic (N--I) 
phase transition. In the case of 5CB, these 
results have led to 
some discussion whether a spinodal dewetting
mechanism driven by van der Waals forces is at work, as proposed by
Vandenbrouck et al. \cite{VanV99}, or wether the
instability is driven by a pseudo--Casimir force based on the
restricted spectrum of director fluctuations in thin nematic films
\cite{ZihP00}. In the present paper, we show 
that neither is true for $n$CB thin films. Instead, the instability 
is caused by textures in the nematic film which largely determine the 
characteristic wavelength of the emerging pattern. This is not the case, 
however, in 5AB$_4$. 
 
5CB and 8CB were obtained from Merck KGaA
(Darmstadt, Germany) and Frinton Laboratories Inc. (Vineland, NJ)
respectively. They were used without further purification. Silicon
wafers (100--oriented, p--(Boron--) doped) with a native oxide layer
of 2 nm provided by Wacker Chemitronics (Burghausen, Germany) were
used as solid substrates. The wafers were cut to samples approximately 
1 cm$^2$ in size
and cleaned with a Snowjet$^{TM}$, which uses a cold CO$_2$ gas stream
mixed with small particles of dry ice to remove particulate and
organic contamination, followed by ultrasonication in ethanol, acetone, 
and hexane, subsequently.  

Immediately after this cleaning process, LC films were
spincast onto the samples from hexane solutions. Variation of
concentration and spinning rate allows to deposit films of variable thickness. The preparation procedure was 
performed in 
a class 100 clean room environment at room temperature. Therefore, the
films were initially in the nematic (5CB) or smectic A (8CB) state, respectively.
Film thicknesses were recorded with an ellipsometer (Optrel GbR,
Berlin, Germany). The samples were placed on a Linkam THMSG 600 heat
stage (temperature control better than 0.1 °C) and observed 
{\it in-situ} with a
Zeiss Axiophot microscope equipped with a digital camera. Unless
otherwise noted, no polarizers were used in the microscope setup.

\begin{figure}
\begin{center}
\includegraphics{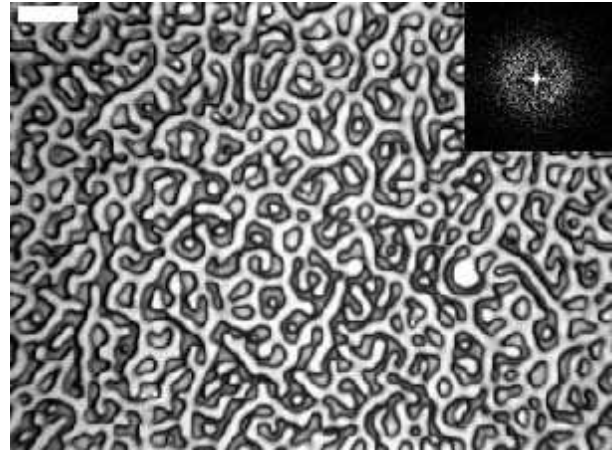}
\end{center}
\caption{\label{fig1}Surface instability of an 85.9(3) nm thick film
  of 8CB at 39.9(1) °C. The scale bar has a length of 100 $\mu$m. The
  inset Fourier transform gives a wavelength of 40 $\mu$m.}
\end{figure}

Observations at room temperature showed films of 5CB and 8CB to be
stable for hours at thicknesses ranging from 50 nm to 200 nm. Close to
the nematic--isotropic transition temperature a surface undulation
with a characteristic wavelength can be observed in both types of samples
 (see figure \ref{fig1}). 
 
\begin{figure}
\begin{center}
\includegraphics{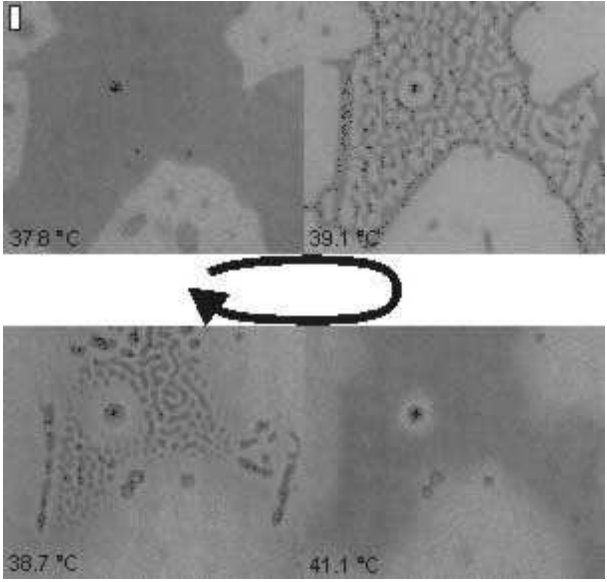}
\end{center}
\caption{\label{fig2} Heating/ cooling sequence showing the transient
  instability in a patch of 8CB (scale bar = 25 $\mu$m).}
\end{figure}

The observed undulation does not lead to a complete dewetting of the
LC film but rather disappears once the phase transition is complete. 
The film becomes homogeneous again in the isotropic phase. Upon cooling, a
similar transient instability occurs. Complete dewetting of the
samples will occur only by heating a few degrees above the N--I
transition, resulting in an array of isotropic droplets. Heating/
cooling of the sample between the nematic and isotropic phase can be
repeated as indicated in fig. \ref{fig2}, every time resulting in nearly
identical undulation patterns. The instability will develop only if
the samples are heated or cooled. Keeping them at a fixed temperature
close to the N--I phase transition will result in the nucleation of
only a few holes in the film. These observations have been made for 
5CB and for 8CB, at all film thicknesses investigated. The pattern was 
found to vanish faster for thicker films after completion of the 
phase transition.
Polarization microscopy confirmed the coincidence of the
instability with the N--I phase transition.  

\begin{figure}
\begin{center}
\includegraphics{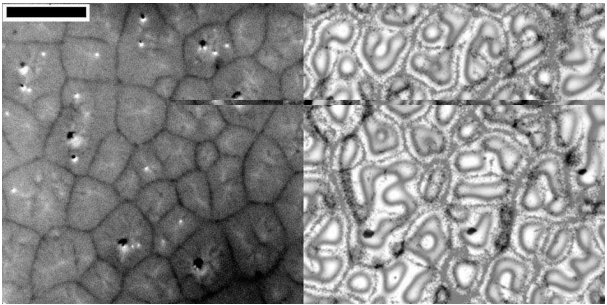}
\end{center}
\caption{\label{fig3}Left: 86 nm thick film of 8CB in the nematic
  state. Borders between nematic domains appear as dark lines. Right:
  Picture of the surface undulation superimposed on the picture of the
  nematic film.  Bright areas (thick film) coincide with domain
  borders in the nematic phase. (scale bar = 100 $\mu$m)}
\end{figure}

Careful examination of the contrast profile of the images revealed the 
formation of nematic domains in 8CB films when heated above the 
smectic A--nematic transition temperature
for the first time after preparation (cf. fig. \ref{fig3}). This
domain pattern is preserved during the subsequent heating and cooling
of the samples which was limited to a few degrees around the N--I
transition temperature. On the left hand side of the figure, contrast is enhanced 
in order to clearly show the domain boundaries, while on the right hand side, 
the undulative instability occurring near the phase transition is superimposed on 
the domain boundary pattern. It is clearly seen from the overlay that the 
undulative pattern is strongly correlated with the domain boundary pattern. 
Since films of 5CB are nematic at room temperature, the domain pattern 
observed immediately after
preparation is less pronounced, but nevertheles present. 
As in the case of 8CB films, this pattern is not
influenced during repeated temperature changes of the sample.

\begin{figure}
\begin{center}
\includegraphics{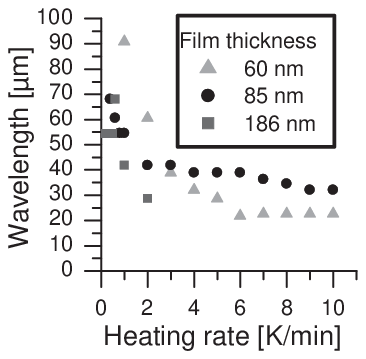} \includegraphics{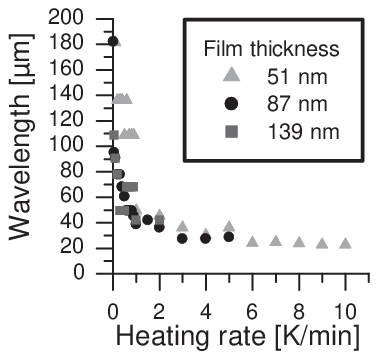}
\end{center}
\caption{\label{fig4}Dependence of undulation wavelength on heating
  rate and film thickness in 8CB (left) and 5CB (right)}
\end{figure}

Since we have found that the undulations appear only when the
sample temperature is swept through the phase transition, it is of 
interest to investigate the impact of the heating rate. We have thus varied 
the heating rate form 0.01 K/min to
10 K/min. As shown in fig.~\ref{fig4}, the heating rate is indeed a major defining
parameter for the wavelength of the undulation: The faster
the samples are heated to the isotropic phase, the smaller is the
undulation wavelength.
 
To explain the instability {\em as such} we propose the following
scenario: After preparation (for 5CB) or after heating to the
nematic phase for the first time (for 8CB), a pattern of nematic domains
exists in the films that is preserved during the course of the
experiment, since close to the substrate nematic or even smectic order
exists even at temperatures substantially above the clearing point
(see e.g. \cite{KOcB00}). The domain boundaries act as nucleation sites for the
isotropic phase upon further heating. Since the surface tension of the
liquid crystalline compounds in question is several percent higher in the
isotropic than in the nematic phase (see e.g. \cite{GanF78}) , a
Marangoni flow drives the instability. Once the phase transition is
completed, the gradient in surface tension vanishes, resulting in a
flat isotropic film. It is possible to explain the pronounced effect 
of the heating upon the
wavelength of the undulation with a few assumptions on the nucleation sites.
This is beyond the scope of this note, and will be published in a forthcoming paper. 
It should be noted that apparently similar observations in 5AB4 \cite{HerJ98} do not belong 
to this class of nucleated undulation phenomena, since in that study the temperature 
was kept constant, and the undulation was not transient, but remained once it was formed. 
  
It is interesting to note that there is no systematic dependence of the patterns 
observed with $n$CB on the film thickness. 
Samples of different thickness (3 thicknesses each for 5CB and 8CB)
show no pronounced dependence of the undulation wavelength on film
thickness, as shown in fig.~\ref{fig4}. In particular, no indication of a 
quadratic dependence of
the wavelength on film thickness could be detected, which would be
expected of a spindoal dewetting scenario, which might be assumed to be present 
on the basis of the appearance of the patterns.

\end{document}